# Phase separation in superconducting and antiferromagnetic $Rb_{0.8}Fe_{1.6}Se_2$ probed by Mössbauer spectroscopy

V. Ksenofontov*[1], G. Wortmann[2], S. Medvedev[1,3],

V. Tsurkan[4,5], J. Deisenhofer[4], A. Loidl[4], and C. Felser[1]

[1]*Institut Anorganische und Analytische Chemie, Johannes Gutenberg-Universität, D-55099 Mainz, Germany*

[2]*Department Physik, Universität Paderborn, D-33095 Paderborn, Germany*

[3]*Max-Planck Institute for Chemistry, D-55128 Mainz, Germany*

[4]*Experimental Physics V, University of Augsburg, D-86159 Augsburg, Germany*

[5]*Institute of Applied Physics, Academy of Sciences, MD-2028, Chişinău, Republic of Moldova*

**Abstract:**

$^{57}$Fe-Mössbauer studies of superconducting $Rb_{0.8}Fe_{1.6}Se_{2.0}$ with $T_C$ = 32.4 K were performed on single-crystalline and polycrystalline samples in the temperature range 4.2 K to 295 K. They reveal the presence of 88% magnetic and 12% non-magnetic $Fe^{2+}$ species with the same polarization dependence of their hyperfine spectra. The magnetic species are attributed to the 16i sites of the √5 x √5 x 1 superstructure, and non-magnetic species to a nano-sized phase postulated from recent studies of superconducting $K_xFe_{2-y}Se_2$ systems. The $^{57}$Fe-spectrum of a single-crystalline sample in an external field of 50 kOe applied parallel to the crystallographic c-axis confirms the antiferromagnetic order between fourfold ferromagnetic Fe(16i) supermoments. The results are discussed with respect to the coexistence of superconductivity and magnetism as well as nano-scale phase separation in $A_xFe_{2-y}Se_2$ systems.

PACS number(s): 74.70.-b, 76.80.+y, 75.30.-m

* v.ksenofontov@uni-mainz.de

The new Fe-based superconductors are in the center of recent interest since superconducting (SC) transition temperatures $T_C$ up to 55 K were observed [1]. These new superconductors all have the same structural motif, namely Fe-X layers, where X is either As or Se, both also modified by substitution. Superconducting FeSe and related Te-substituted systems possess the simplest structure, consisting of Fe-Se layers only. FeSe exhibits a relatively low $T_C$ value of 8.5 K, however, a significant increase of $T_C$ up to 36.7 K is observed under pressure of ~9 GPa [2-4]. At present, the microscopic mechanisms of superconductivity in these Fe-based superconductors are widely discussed, e.g. as based on spin fluctuations or complex interactions between occupied and non-occupied structures at the Fermi-edge of the valence band, built up dominantly from Fe 3d-band electrons [5]. The observation of superconductivity in magnetic members of the $SrFe_2As_2$ called into question the old paradigm that superconductivity cannot coexist with magnetism [5], similar to previous observations in heavy-fermion superconductors [6]. This actual discussion was stimulated by the reports of coexistence of superconductivity and magnetism in a new family of Fe-based superconductors, namely the $A_xFe_{2-y}Se_2$ (A = K, Rb, Cs) systems exhibiting $T_C$ values up to 32 K and antiferromagnetic ordering temperatures above 500 K, both properties representing bulk samples [7,8]. These novel $A_xFe_{2-y}Se_2$ compounds are closely related to superconducting, but non-magnetic FeSe, but now with the Fe-Se layers separated by alkali ions. In this respect they are similar to the above mentioned $SrFe_2As_2$ systems with $ThCr_2Si_2$ structure, however, with an important difference by a $\sqrt{5}$ x $\sqrt{5}$ x 1 superstructure modulating the Fe deficiency in the Fe-Se layers by the formation of Fe ions in 16i positions and of empty Fe sites in 4d positions [9,10]. Very recent reports about a nano-sized filamentary phase formed concomitantly with the main $\sqrt{5}$ x $\sqrt{5}$ x 1 superstructure have provided new aspects about coexistence of superconductivity and magnetism in $K_xFe_{2-y}Se_2$ systems. Yuan *et al.* [11] reported on a nano-scale phase separation of the magnetic $\sqrt{5}$ x $\sqrt{5}$ x 1 superstructure together with a minority K-deficient superconducting phase in $K_{0.75}Fe_{1.75}Se_2$. Ricci *et al.* [12] reported also on a nano-scale phase separation in $K_{0.8}Fe_{1.6}Se_2$ and suggested that the coexistence of insulating vacancy-ordered magnetic domains with $\sqrt{5}$ x $\sqrt{5}$ x 1 structure and metallic non-magnetic domains is responsible for superconductivity.

In this study we provide strong evidence for this phase separation in a well characterized superconducting single-crystalline $Rb_{0.8}Fe_{1.6}Se_{2.0}$ sample investigated by $^{57}$Fe-Mössbauer spectroscopy, which is especially suited because of the site specific information on the local magnetic, electronic and structural properties of the Fe species, being the active players for the magnetic and superconducting properties.

Single crystals of the Rb-Fe-Se system were grown by Bridgman method using as starting materials the elemental Rb (99.75%) and high-purity FeSe previously synthesized by solid state reactions [13]. Platelet-like single crystals with diameters of ~10 mm were separated from the solidified melt. XRD studies at 295 K on a powdered sample, grinded from small crystals under inert conditions, revealed lattice parameters a = 3.9228(7) Å and c = 14.5909(38) Å assuming I4/mmm symmetry. The composition of the sample was determined by wave length-dispersive electron-probe microanalysis by averaging over 30 spots on the surface of a freshly cleaved single-crystalline sample as $Rb_{0.80(3)}Fe_{1.60(1)}Se_{2.00(4)}$, named in the following in short $Rb_{0.8}Fe_{1.6}Se_2$. For all details about sample preparation and characterization see Ref. 13.

Magnetic susceptibility measurements delivering the superconducting and magnetic properties were performed in the temperature range 2 – 600 K and in magnetic fields up to 50 kOe using a SQUID magnetometer [13]. $^{57}$Fe-Mössbauer studies were recorded using a $^{57}$Co(Rh) source with a constant-acceleration spectrometer. Low-temperature spectra were taken in transmission geometry using conventional bath cryostat as well as a cryostat equipped with a superconducting solenoid. The evaluation of the spectra with complex hyperfine interactions was performed with the RECOIL 1.03 fit routine [14].

Two different absorbers were prepared from the single crystals of $Rb_{0.8}Fe_{1.6}Se_{2.0}$ for the Mössbauer studies. A powdered sample was prepared by crushing small pieces of the single crystals under strictly inert conditions in a glove box and compressing the powder between plastic windows of the absorber holder. This yields a textured sample with the crystalline c-axis oriented preferentially perpendicular to the absorber plane as evidenced from the $^{57}$Fe-spectra. Another absorber was prepared by attaching thin single-crystalline flakes, separated from the bulk single crystals by the so-called "scotch-tape" technique. Mössbauer spectra of this sample reflect the quasi single-crystalline quality of the absorber, called in the following "mosaic" absorber, with the *c*-axis oriented perpendicular to the absorber plane and the *a,b*-axes randomly within the absorber plane, which resulted in a transmission of the gamma rays parallel to the c-axis within a range of ± 5°.

Fig. 1 exhibits the temperature dependence of zero-field cooled (ZFC) and field-cooled (FC) susceptibilities of a single-crystalline $Rb_{0.8}Fe_{1.6}Se_{2.0}$ sample measured in a field of 10 Oe applied along the *c*-axis. The transition to a fully superconducting state occurs at $T_C$ = 32.4 K within 1.5 K. The bulk nature of superconductivity was concluded from specific heat measurements yielding a volume fraction above 90% [13]. The inset in the Fig. 1 shows the magnetic hysteresis measured at 2 K after cooling in zero field. The width of the diamagnetic response corresponds to a critical current density of $1.6 \times 10^4$ A/cm$^2$ [13]. Finally, from resistivity studies upper critical fields $H_{c2}$ = 25 T with the magnetic field parallel to the *c*-axis and $H_{c2}$ = 63.3 T for the perpendicular direction were derived [13]. These latter values are lower than the ones reported in other studies of $Rb_{1-x}Fe_{2-y}Se_2$, while the other superconducting properties compare well with those reported [8, 15].

Mössbauer spectra of the powdered sample were measured first at 4.2 K (Fig. 2a) and then with increasing temperature up to room temperature. All spectra are very similar to that observed at 4.2 K, as long as the absorber was kept under inert conditions. They consist of a dominant magnetic sextet with 88(1)% spectral intensity of the Fe sites and of a non-magnetic quadrupole doublet with 12(1)% intensity. Both spectra show clear evidence for a strong texture by a deviation of the line intensity ratios of the magnetic subspectra from 3:2:1:1:2:3 and of the quadrupole doublet from a 1:1 ratio as expected for a fully polycrystalline absorber. Fit analysis of the dominant magnetic subspectrum in Fig. 2a was only successful when a combined magnetic-dipole and electric-quadrupole interaction was applied, where the direction of the magnetic hyperfine field does not coincide with the main axis of the electric-field gradient (EFG) tensor, including the asymmetry parameter η for the EFG. As described in [9,10] the formation of the superstructure of $A_xFe_{2-y}Se_2$ by an ordering of the vacant Fe sites is connected with a first-order structural phase transition, triggered by the concomitant onset of strong magnetic interactions within and between the supermoments formed by 4 Fe neighbors with parallel spins, coupled antiferromagnetically with the neighboring supermoments of Fe quartets. The point symmetry of the magnetic 16i sites is lowered by this structural phase transition from *4m* to *1*, as reflected by the changes of the Fe-Se and Fe-Fe distances as well as Se-Fe-Se binding angles, analyzed for the magnetic $K_{0.8+x}Fe_{1.6-y}Se_2$ system [16]. The derived EFG tensor at the 16i sites, $\Delta E_Q$ = +1.2(1) mm/s with η = 0.12(2) and an angle θ = 45(2)° between the main axis of the EFG and the magnetic hyperfine field, $B_{hf}$ = 283.0(3) kOe at 4.2 K, is drastically different in magnitude and direction to that in the undistorted Fe-Se layers of superconducting $Fe_{1.01}Se$ with $\Delta E_Q$ = -0.32 mm/s at 4.2 K [3,4]. This large value of $\Delta E_Q$ can only originate from the local $Fe^{2+}$ moment in a high-spin state, thereby providing a large orbital contribution to the EFG. From the relative intensity ratios of the magnetic 6-line pattern a preferred orientation of the magnetic hyperfine fields, and therewith of the crystalline *c*-axes, with an averaged angle of 40(2)° with respect to the direction of the γ-rays, is derived. The non-magnetic site could be well adjusted by a quadrupole splitting with the same polarization dependence and $\Delta E_Q$ = -0.32 mm/s, which in sign and magnitude is identical to that observed in superconducting $Fe_{1.01}Se$ with the main axis of the EFG parallel to the *c*-axis [3,4]. This is an important information showing that these non-magnetic Fe sites are located either on the formally vacant 4d sites of the √5 x √5 x 1 superstructure of $Rb_{0.8}Fe_{1.6}Se_2$ [9,10] or in the above mentioned nano-sized phase with √2 x √2 x 1 structure. It is important to note that the orientation of the c-axes of both phases coincide, as evidenced in the superstructure reflections presented for $K_xFe_{2-y}Se_2$ [11,12].

The spectrum of the mosaic sample shown in Fig. 2b provides direct information about the high degree of orientation the c-axes of the magnetic phase and therewith of the magnetic hyperfine fields $B_{hf}$ of the Fe 16i-sites, evidenced by the strongly reduced intensities of the lines 2 and 5. The magnetic subspectrum, again with 88% relative intensity, is well reproduced by the fit with the parameters given above for the powdered sample, providing also the unusual intensity ratios of the magnetic sextet, amounting to 2.86:0.30:1.04:0.96:0.10:3.14 (see Fig. 2b, from the left to the right). This indicates a pronounced deviation from the above mentioned intensity ratio of purely magnetic sextets, which is caused by complex non-collinear hyperfine interactions and most pronounced for the present quasi single-crystalline absorber, oriented with an averaged angle of 6(2)° between the direction of the γ-rays with respect to $B_{hf}$ and, correspondingly, to the crystallographic *c*-axis. This information agrees well with the polarization dependence of the quadrupole doublet of the non-magnetic sites, being now very close to 3:1, resulting in an angle of 8(3)° between the gamma rays and the main axis of the EFG, being therewith parallel to the *c*-axes of the mosaic sample.

Here we shortly comment on two $^{57}$Fe-Mössbauer studies performed on $K_xFe_{2-y}Se_2$ systems. $K_{0.80}Fe_{1.76}Se_{2.00}$ was studied in the wide temperature range from 10 K up to and above the Neel temperature $T_N$ = 532 K by Ryan *et al*. [17], $K_{0.86}Fe_{1.73}Se_2$ was studied between 16 K and 295 K by Li *et al*. [18]. In both studies, the $^{57}$Fe-spectra of mosaic absorbers prepared from single-crystalline flakes as in the present study, exhibited for the magnetic Fe sites the same spectral features as for the mosaic $Rb_{0.8}Fe_{1.6}Se_2$ sample shown in Fig. 2b, e.g. the same unusual intensity ratios of the dominant magnetic sextet pointing to the same spectral parameters for the Fe moments oriented parallel to the crystal c-axis. However, in both studies obviously a collinear quadrupole interaction was adjusted to the magnetic sites, quoted as $\Delta E_Q$ = +0.33(2) mm/s in [17] and with a similar value in [18], not quoted but extracted from an inspection of the fitted subspectra. By this fitting both studies could not account for the appearance of "anomalous" line intensities attributed to components 2 and 5 as well as for clearly visible asymmetries in line pairs 1-6 and 3-4. In addition, the line positions of the line pairs 2-5 and 3-4 could not be correctly adjusted. All these properties are well described in the present study with much higher statistical accuracy of the spectra than in [17,18]. It is important that for $K_{0.80}Fe_{1.76}Se_{2.00}$ also a non-magnetic component with a relative intensity of 12(2)% intensity, identical, also in position, to the present spectrum in Fig.2b, was observed [17], however the quadrupole splitting of this site was not analyzed because of the lower spectral resolution. The $^{57}$Fe-spectrum of the mosaic absorber of $K_{0.86}Fe_{1.73}Se_2$ [18] exhibits a similar non-magnetic component, but with a higher intensity of ~25%, while a powdered absorber exhibits a non-magnetic component of even higher intensity [18]. We interpret this fact, at least in part, from decomposition products, similar to the observations in FeSe [4]. We will discuss these observation in details as well as the temperature dependence of the hyperfine parameters and spectral areas in a forthcoming publication [19].

Using the advantage of high degree of orientation of the crystalline *c*-axes of the mosaic $Rb_{0.8}Fe_{1.6}Se_2$ sample, an additional $^{57}$Fe-spectrum was measured at 4.2 K in an external magnetic field of $B_{ext}$ = 50 kOe applied parallel to the *c*-axes and to the direction of the gamma rays (Fig. 2c).

The spectral features are now drastically changed. The magnetic sites are split into two components of exact equal intensity with magnetic hyperfine fields of $B_{hf}(1) = B_{hf} + B_{ext}$ and $B_{hf}(2) = B_{hf} - B_{ext}$, with $B_{hf}$ = -283.0(3) kOe. Therefore $B_{hf}(1)$ is attributed to Fe sites building up the supermoments with their combined moments arranged parallel to $B_{ext}$, as demonstrated by the observed value, $B_{hf}(1)$ = -232.9(5) kOe, which is exactly the sum of $B_{hf} + B_{ext}$. Accordingly, $B_{hf}(2) = B_{hf} - B_{ext}$ = -333.0(5) kOe is monitored by the Fe sites with their moments arranged antiparallel to $B_{ext}$. The non-magnetic Fe sites experience the same external field, their quadrupole spectrum is now split by $B_{ext}$ parallel to $V_{zz}$, the main component of the EFG, as shown in Fig. 2c, with a fitted value of $B_{hf}$ = 50.1(5) kOe, exactly the value of $B_{ext}$.

Since the external field was applied parallel the c-axis of the oriented crystals of $Rb_{0.8}Fe_{1.6}Se_2$, the above observations prove the antiferromagnetic arrangement of the magnetic Fe-sites in their fourfold supermoments arranged parallel or anti-parallel to the *c*-axis as derived from neutron scattering studies for various $A_xFe_{1-y}Se_2$ systems [9, 20]. The narrow line widths of both magnetic and non-magnetic Fe sites in all spectral features point to a rather perfect composition of the present $Rb_{0.8}Fe_{1.6}Se_2$ sample. Any kind of Fe off-stoichiometry at the 16i sites, e.g. only 3 Fe forming a ferromagnetic supermoment, should be reflected in the line width of the 16i-sites or by additional magnetic sextets in the $^{57}$Fe-spectra with different hyperfine parameters, which is not the case. The present data disprove a possible arrangement with two different Fe moments, as discussed for the $Rb_{0.8}Fe_{1.6}Se_2$ system by Pomjakushin *et al.* [20]. The latter case should be reflected by two different magnetic subspectra in Fig.2b and four magnetic subspectra in Fig.2c, which are not observed.

One of the most important present finding is that 12% of the $Fe^{2+}$ ions are occupying a well defined non-magnetic site, following in its polarization dependence the principal orientation the c-axis of the √5 x √5 x 1 superstructure. The following discussion is devoted to attribute these non-magnetic Fe sites either (i) to the formally empty 4d-sites of the √5 x √5 x 1 superstructure or (ii) to the non-magnetic minority phase, observed in $A_xFe_{1-y}Se_2$ superconductors as mentioned above.

(i) Considering an exact stoichiometry of the present sample as $Rb_{0.8}Fe_{1.6}Se_2$ and assuming a full occupancy of the 16i sites, as evidenced by the high spectral quality of the magnetic subspectra, there is no Fe left to occupy the "empty" 4d sites in the √5 x √5 x 1 superstructure. The actually observed 12% spectral intensity of non-magnetic sites, assuming again a full occupancy of the 16i sites, would correspond to a ~55% occupation of the 4d sites. This is a very high value which contradicts the so-called vacancy order of Fe 4d-sites, which is considered as a crucial factor for the formation of the magnetic superstructure.

From XRD analysis for superconducting $K_{0.737}Fe_{y1.613}Se_2$ and $K_{0.775}Fe_{1.631}Se_2$, occupancy of the 4d-sites by 3.2% and by 7.8%, respectively, was reported together with a full occupation of the 16i-sites [10]. This contrasts with non-superconducting $K_{0.8623)}Fe_{1.563(4)}Se_2$, where an 4d-site occupancy as high as 22%, connected with 92% occupation of the 16i-sites, was reported [16]. Summarizing, one can conclude that 55% occupation of the 4d sites is not compatible with XRD analysis of super-conducting $A_xFe_{2-y}Se_2$ systems as well as with the perfect magnetic properties of the 16i-sites in the √5 x √5 x 1 superstructure evidenced in the present $^{57}Fe$ spectra.

(ii) Most important for the attribution of the non-magnetic Fe sites to the nano-sized minority phase is the fact that the principal c-axes of both majority and minority phases coincide. Now we can assign the observed 88% magnetic Fe sites in the $^{57}Fe$-spectra to the 16i-sites of the √5 x √5 x 1 phase and the 12% non-magnetic Fe sites to this minority phase, attributed in [11,12] to the superconducting properties of the whole sample. The identical direction of the c-axes of both phases explains immediately the observed polarization dependence of the quadrupole interaction of the non-magnetic Fe sites, which is in addition identical in magnitude and polarization dependence to that observed in superconducting FeSe.

Finally we want to point out that it emerges from the numerous reports on the $A_xFe_{2-y}Se_2$ systems that a relative Fe amount of 1.60, within experimental uncertainty, marks the lower borderline for the occurrence of superconductivity. For instance, M. Wang *et al.* [21] reported that $Rb_{0.89}Fe_{1.58}Se_2$ is non-superconducting but exhibits the √5 x √5 x 1 structure and strong magnetism with $T_N$ = 475 K. We observed for a non-superconducting $Rb_{0.79}Fe_{1.49}Se_2$ sample magnetic $^{57}Fe$ spectra with identical hyperfine parameters to the present sample, but without non-magnetic sites [19]. All these findings support the location of the non-magnetic sites in the nano-scaled minority phase.

Concluding, we have shown that $^{57}Fe$-Mössbauer spectra of powdered and single-crystalline samples of $Rb_{0.8}Fe_{1.6}Se_2$ exhibit well-resolved hyperfine spectra with a dominant magnetic site and a non-magnetic site in an intensity ratio of 88(1)% to 12(1)%. The magnetic sites can by perfectly adjusted by a fit analysis of $Fe^{2+}$ ions with a non-collinear magnetic-dipole and electric quadrupole interactions. The spectra of the single-crystalline absorber prove the orientation of the moments parallel to the crystalline c-axis. The spectra in an external magnetic field prove the antiferromagnetic arrangement of ferromagnetically coupled Fe supermoments within the √5 x √5 x 1 superstructure.

The non-magnetic Fe sites exhibit a quadrupole splitting with the same polarization dependence as the magnetic sites, indicating that they are located either on the formally vacant 4d-sites of the dominant √5 x √5 x 1 superstructure or in a newly detected nano-sized filamentary phase co-existing with the dominant magnetic phase. The present data favor the latter case, thereby supporting the findings from other $A_xFe_{2-y}Se_2$ systems that this non-magnetic phase, with similar spectral properties as observed in superconducting FeSe, is also responsible for superconductivity in $Rb_{0.8}Fe_{1.6}Se_2$.

Acknowledgement: This work was supported by the DFG within the SPP 1458 by the grants FE 633/10-1, ER 539/6-1 (Mainz) and DE1762/1-1 (Augsburg) and by the TTR80 (Augsburg-Munich).

**Figures**

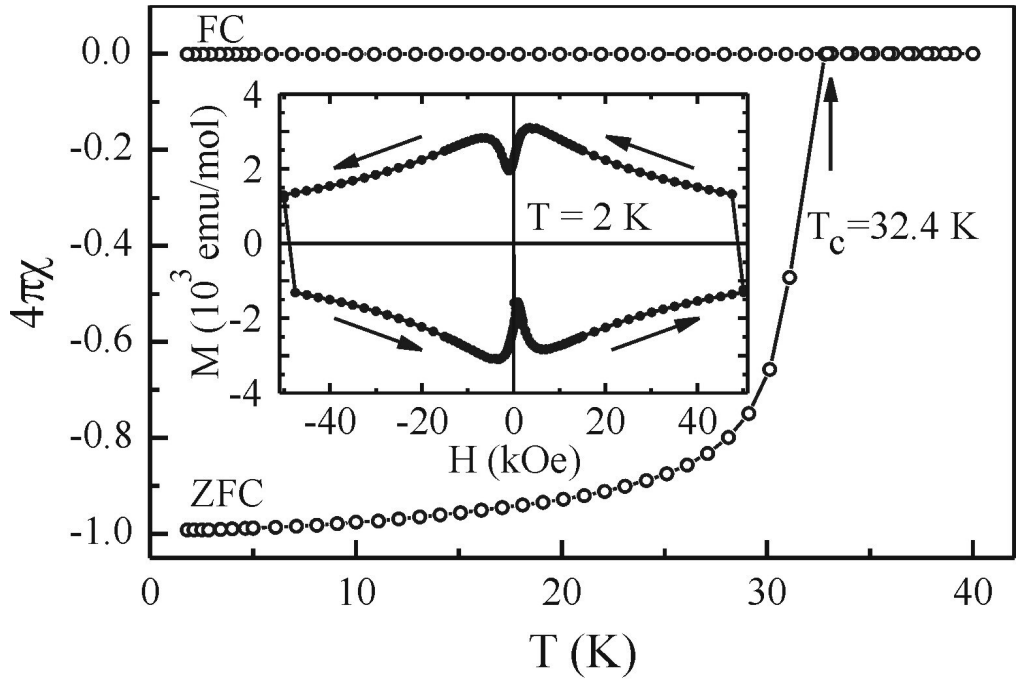

Fig. 1

Temperature dependence of the susceptibility of a single crystalline sample $Rb_{0.8}Fe_{1.6}Se_2$ measured in zero-filed cooled (ZFC) and in magnetic field of 10 Oe applied along the *c*-axis (FC). Inset shows the magnetization hysteresis loop measured at T = 2 K.

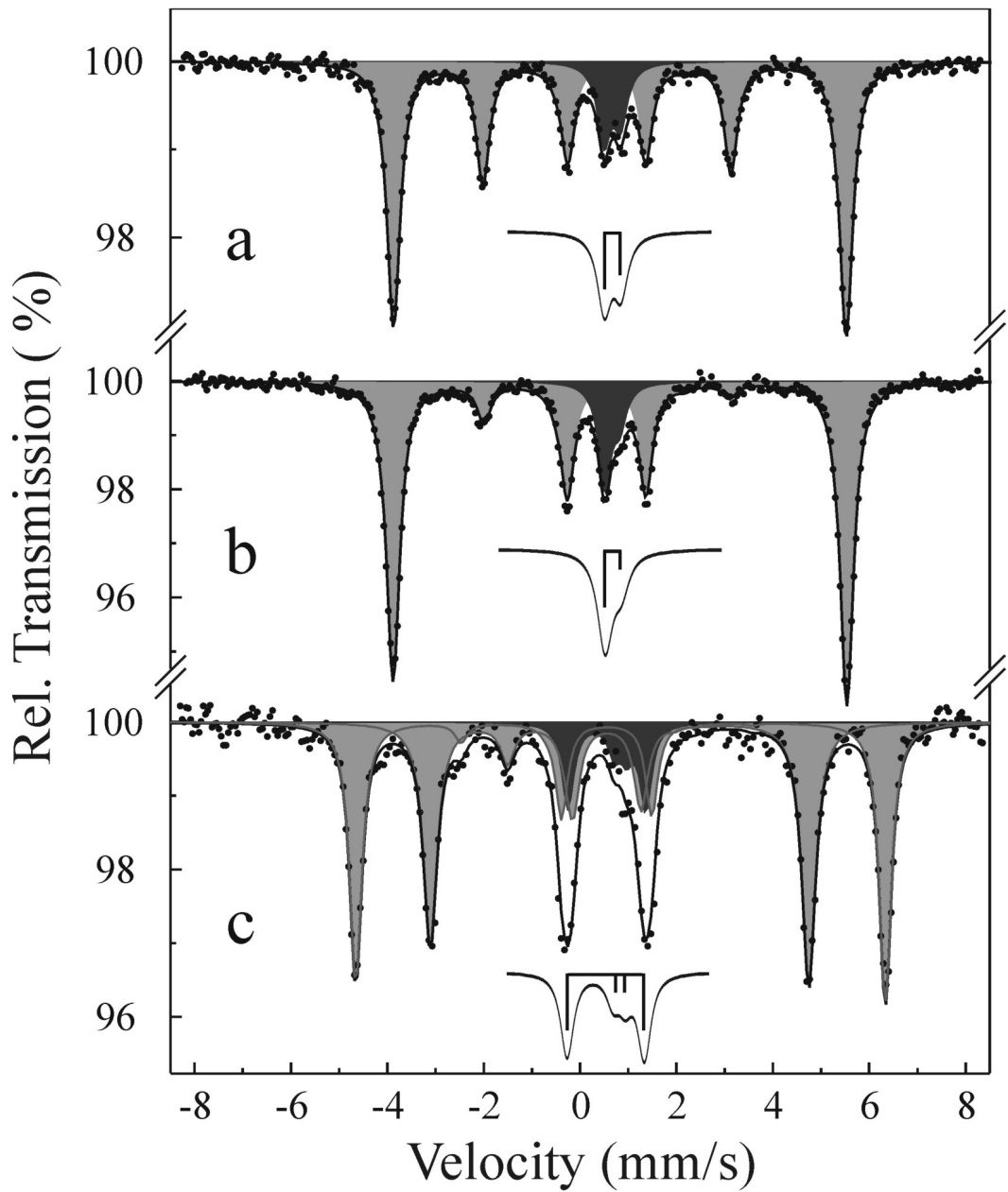

Fig. 2: $^{57}$Fe-Mössbauer spectra of $Rb_{0.8}Fe_{1.6}Se_2$ measured at T = 4.2 K of
a) powdered sample. b) mosaic sample with the c-axis parallel to the direction of the gamma rays. c) mosaic sample as in b) with external magnetic field of 50 kOe applied parallel to the c-axis. Subspectra of the magnetic Fe sites are marked in dark grey, subspectra of non-magnetic Fe sites are shown in black. Insets indicate in a) and b) the quadrupole–split subspectra of the non-magnetic Fe sites with different polarizations and in c) the quadrupole-split non-magnetic Fe sites experiencing the external field. The velocity scale is given with respect to α-Fe at 295 K.